\documentstyle[12pt]{article}
\input html.sty
\input epsf
\begin{document}
\title{MATTERS OF GRAVITY, The newsletter of the APS Topical Group on 
Gravitation}
\begin{center}
{ \Large {\bf MATTERS OF GRAVITY}}\\
\bigskip
\hrule
\medskip
{The newsletter of the Topical Group on Gravitation of the American Physical 
Society}\\
\medskip
{\bf Number 17 \hfill Spring 2001}
\end{center}
\begin{flushleft}

\tableofcontents
\vfill
\section*{\noindent  Editor\hfill}

\medskip
Jorge Pullin\\
\smallskip
Center for Gravitational Physics and Geometry\\
The Pennsylvania State University\\
University Park, PA 16802-6300\\
Fax: (814)863-9608\\
Phone (814)863-9597\\
Internet: 
\htmladdnormallink{\protect {\tt{pullin@phys.psu.edu}}}
{mailto:pullin@phys.psu.edu}\\
WWW: \htmladdnormallink{\protect {\tt{http://www.phys.psu.edu/\~{}pullin}}}
{http://www.phys.psu.edu/\~{}pullin}\\
\hfill ISSN: 1527-3431
\begin{rawhtml}
<P>
<BR><HR><P>
\end{rawhtml}
\end{flushleft}
\pagebreak
\section*{Editorial}

Not much to report here. If you are burning to have Matters of 
Gravity with you all the time, the newsletter is now available for
Palm Pilots, Palm PC's and web-enabled cell phones as an
Avantgo channel. Check out 
\htmladdnormallink{\protect {\tt{http://www.avantgo.com}}}
{http://www.avantgo.com} under technology$\rightarrow$science.
The next newsletter is due September 1st.
If everything goes well this newsletter should be available in the
gr-qc Los Alamos archives 
(\htmladdnormallink{{\tt http://xxx.lanl.gov}}{http://xxx.lanl.gov})
under number gr-qc/0102044. To retrieve it
send email to
\htmladdnormallink{gr-qc@xxx.lanl.gov}{mailto:gr-qc@xxx.lanl.gov}
with Subject: get 0102044
(numbers 2-16 are also available in gr-qc). All issues are available in the
WWW:\\\htmladdnormallink{\protect {\tt{http://gravity.phys.psu.edu/mog.html}}}
{http://gravity.phys.psu.edu/mog.html}\\ 
A hardcopy of the newsletter is
distributed free of charge to the members of the APS
Topical Group on Gravitation upon request (the default distribution form is
via the web) to the secretary of the Topical Group. 
It is considered a lack of etiquette to
ask me to mail you hard copies of the newsletter unless you have
exhausted all your resources to get your copy otherwise.
\par
If you have comments/questions/complaints about the newsletter email
me. Have fun.
\bigbreak

\hfill Jorge Pullin\vspace{-0.8cm}
\section*{Correspondents}
\begin{itemize}
\item John Friedman and Kip Thorne: Relativistic Astrophysics,
\item Raymond Laflamme: Quantum Cosmology and Related Topics
\item Gary Horowitz: Interface with Mathematical High Energy Physics and
String Theory
\item Richard Isaacson: News from NSF
\item Richard Matzner: Numerical Relativity
\item Abhay Ashtekar and Ted Newman: Mathematical Relativity
\item Bernie Schutz: News From Europe
\item Lee Smolin: Quantum Gravity
\item Cliff Will: Confrontation of Theory with Experiment
\item Peter Bender: Space Experiments
\item Riley Newman: Laboratory Experiments
\item Warren Johnson: Resonant Mass Gravitational Wave Detectors
\item Stan Whitcomb: LIGO Project
\end{itemize}
\vfill
\pagebreak

\section*{\centerline {
APS Prize on gravitation:}}
\addtocontents{toc}{\protect\medskip}
\addtocontents{toc}{\bf APS TGG News:}
\addtocontents{toc}{\protect\medskip}
\addcontentsline{toc}{subsubsection}{\it  
APS Prize on gravitation, by Clifford Will}
\begin{center}
    Clifford Will, Washington University St. Louis
\htmladdnormallink{cmw@howdy.wustl.edu}
{mailto:cmw@howdy.wustl.edu}
\end{center}
The Topical Group on Gravitation and the APS have established a
Prize in Gravitational Physics, and have begun a campaign to raise \$200,000
to endow the prize.  Through the generosity of Dr. David Lee, a 1974 Caltech
Ph.D. in gravitational physics, a challenge gift of up to \$100,000 has been
promised, to match every dollar raised from other sources.

\medskip
The prize was established to recognize outstanding achievements in
gravitational physics, both theoretical and experimental.  The APS plans to
name it the {\it Einstein Prize in Gravitational Physics}.
As of January 2001, we have raised \$100,000 (\$50,000 from Dr. Lee,
\$50,000 from TGG members).  

\medskip
\centerline
{{\it Please give generously to support this new TGG Prize!}}

\medskip
\noindent
You may make a lump sum contribution, or pledge an amount to be spread
over some years.  The APS will send a reminder when each installment
is due.
Contributions are tax deductible as a charitable donation, and can be sent
to

\medskip
Gravitational Physics Prize
c/o Darlene Logan, Director of Development
American Physical Society
One Physics Ellipse
College Park MD 20740-3844

\section*{\centerline {
TGG Elections}}
\addcontentsline{toc}{subsubsection}{\it  
TGG Elections, by David Garfinkle}
\begin{center}
    David Garfinkle, Oakland University
\htmladdnormallink{garfinkl@oakland.edu}
{mailto:garfinkl@oakland.edu}
\end{center}
The nominating committee of the Topical Group on
Gravitation has put together the following slate of candidates.

 For Vice Chair: 
    John Friedman, U. of Wisconsin, Milwaukee, 
    Bill Hamilton, Louisiana State U.

 For Members-at-large:  
     Ted Jacobson, U. of Maryland, 
     Pablo Laguna, Penn State, 
     Don Marolf,  Syracuse U.,
     Jennie Traschen, U. Mass. Amherst

      The purpose of this message is to ask for any nominations from the
general membership of the Topical Group on Gravitation.  If any member
is nominated by at least 5 percent of the membership of the topical
group then that member will be added to the ballot.  Please send your
nominations to me at 
\htmladdnormallink{
garfinkl@oakland.edu}{garfinkl@oakland.edu} and give the name of the
person you are nominating and the office (Vice Chair or
Member-at-large).  Nominations must be received by Feb. 15.

\section*{\centerline {
We hear that...}}
\addcontentsline{toc}{subsubsection}{\it  
We hear that, by Jorge Pullin}
\begin{center}
    Jorge Pullin, Penn State
\htmladdnormallink{pullin@phys.psu.edu}
{mailto:pullin@phys.psu.edu}
\end{center}
TGG members James Isenberg, James Hough, and William Unruh have
recently been selected to be APS Fellows. Congratulations!

\vfill\eject

\section*{\centerline {
Experimental Unruh Radiation?}}
\addtocontents{toc}{\protect\medskip}
\addtocontents{toc}{\bf Research Briefs:}
\addtocontents{toc}{\protect\medskip}
\addcontentsline{toc}{subsubsection}{\it  
Experimental Unruh Radiation?, by Matt Visser}
\begin{center}
Matt Visser, Washington University St. Louis
\htmladdnormallink{visser@kiwi.wustl.edu}
{mailto:visser@kiwi.wustl.edu}
\end{center}

Experimental detection of Hawking radiation from real
general-relativistic black holes seems a close to hopeless
proposition. Even the detection of the Hawking radiation that is
expected to arise from condensed matter analog models for general
relativity, while much more accessible than that from true
gravitational black holes, is also currently far from laboratory
realization. Given this, perhaps the next best thing to do is to
attempt an experimental verification of the existence of Unruh
radiation. This is the hope of Pisin Chen (Stanford) and Toshi Tajima
(Austin) who have analyzed the possibility of using intense lasers to
accelerate electrons extremely rapidly [1].

Recall that the Unruh effect implies that a uniformly accelerating
particle will find itself surrounded by a thermal heat bath of
temperature
\[
k T = {\hbar a \over 2\pi c}. 
\hfill\qquad\qquad\qquad (1)
\]
More precisely, uniform acceleration through the usual quantum vacuum
(Minkowski vacuum) of the electromagnetic field will distort the
two-point function of the zero-point fluctuations (ZPF) in such a way
that
\[
\langle E_{i}(-\tau/2) E_{j}(+\tau/2) \rangle = 
{4\hbar\over\pi c^3} \; \delta_{ij} \; 
{(a/c)^4 \over \sinh^4(a \tau/2c)}.
\hfill\qquad\qquad  (2)
\]
Here $\tau$ is the proper time as measured at some fixed position in
the accelerated frame, while $a$ is the acceleration.  As $a \to 0$
\[
\langle E_{i}(-\tau/2) E_{j}(+\tau/2) \rangle = 
{64\hbar c\over\pi  \tau^4},
\]
which recovers the usual unaccelerated Minkowski space result.

In the setup considered by Chen and Tajima [1], they use a
laser-driven classical EM field to accelerate a single
electron. Because they are not accelerating the entire detector, just
a single electron, searching for an Unruh temperature as in equation
(1) is meaningless. Instead, they suggest looking for the effects due
to equation (2): The acceleration of the electron through the
Minkowski vacuum state modifies the correlations in the zero-point
fluctuations of the vacuum, which causes an additional jitter in the
electron's motion, which then modifies the radiation emitted by the
electron --- over and above the classical Larmor radiation. This
additional acceleration-related radiation has a characteristic
acceleration dependence (a distorted thermal spectrum) and a
characteristic angular dependence, which should in principle be
measurable in the not too distant future. In particular there is a
"blind spot" in the angular dependence of the classical Larmor
radiation [1,2], and if you sit in this blind spot any radiation you
see should be traceable to this distortion of the zero-point
fluctuations.

There are two tricky points to keep in mind, one of physics and one of
sociology/linguistics:

(1) There is a maximum electric field beyond which the QED vacuum
falls apart due to copious production of electron-positron pairs
(Schwinger effect). This vacuum breakdown occurs for
\[
{e} E_{max} \approx {m_e c^2\over\lambda_e} = 
{m_e c^2\over\hbar/ (m_e c)} = m_e^2 c^3/\hbar,
\]
and corresponds to a maximum acceleration
\[
{a}_{max} \approx m_e c^3/\hbar \approx 10^{29} \; m/s^2 
\approx 10^{28} \; g_{earth}.
\]
The accelerations posited by Chen and Tajima are up to $10^{25}$
$g_{earth}$, so they are approaching but not quite over this vacuum
breakdown limit. Thus if you succeed in building the experiment
suggested by Chen and Tajima you are close to ultimate limits on this
type of experiment --- there's not much extra maneuvering room.

(2) The linguistic problem is this: If you ultimately succeed in
seeing this ZPF-induced modification to Larmor radiation, should you
really call it the Unruh effect? [2,3] Or should you just call it
basic quantum field theory? After all you are not directly measuring
the Unruh temperature itself.  [To add to the confusion there is a
subspecies of physicist that still does not believe in quantum field
theory (QFT), and instead goes through quite contorted gymnastics to
try to interpret all of quantum physics in terms of a classical
stochastic background of zero-point fluctuations.  I do not expect
this subspecies to be convinced by the experiment, regardless of the
outcome.]

I think it fair to say that most of the relativity and quantum
communities would view a successful experiment along these lines as a
beautiful verification of the basic ideas of flat-space QFT. The
connection with curved-space QFT is tenuous at best, but this does not
reduce the interest in performing this type of experiment.

{\bf References}

\noindent
[1] Testing Unruh Radiation with Ultra-intense Lasers.  Pisin Chen and
Toshi Tajima. Physical Review Letters 83, 256-259 (1999).

\noindent
[2] Blind spot may reveal vacuum radiation. Haret Rosu, 
Physics World, October 1999, 21-22.

\noindent
[3] On the estimates to Measure Hawking Effect and Unruh Effect in the
Laboratory. Haret Rosu, International Journal of Modern Physics D3,
545 (1994); 
\htmladdnormallink{gr-qc/9605032}{http://xxx.lanl.gov/abs/gr-qc/9605032}.

\vfill\eject
\section*{\centerline {Why is the Universe accelerating?}}
\addcontentsline{toc}{subsubsection}{\it  
Why is the Universe accelerating?, by Beverly Berger}
\begin{center}
Beverly Berger, Oakland University
\htmladdnormallink{berger@oakland.edu}
{mailto:berger@oakland.edu}
\end{center}

One of the most important discoveries of the late 20th century was the
evidence from Type Ia supernovae (SNe-Ia) that the expansion of the
Universe is accelerating [1]. If this result holds up, it will have
fundamental significance for gravitation and cosmology. 

Of course, the historically honored mechanism for the gravitational
repulsion needed to provide the acceleration is the cosmological
constant, $\Lambda$, originally proposed (and then retracted) by
Einstein. A competing explanation, originally proposed by Caldwell {\it et
al.} [2], is quintessence, a mechanism to obtain a time dependent
cosmological constant with a scalar field $\phi_Q$ and a potential
$V(\phi_Q)$. However, quintessence models have a large number of
adjustable parameters and are {\it ad hoc} in the sense that there is no
underlying quantum theory for $\phi_Q$. Both the cosmological constant
and quintessence are added to ``standard'' cold dark matter (CDM)
Friedmann-Robertson-Walker (FRW) cosmologies. With suitable adjustments
of parameters, both $\Lambda$CDM and QCDM models can fit the SNe-Ia
data. 

Another important and recently discovered constraint on cosmological
models is the angular dependence of the cosmic microwave background
fluctuations (CMBF) initially detected by COBE and more accurately
extended to smaller angular scales by BOOMERANG and MAXIMA [3]. Assuming
the scale invariant spectrum of initial adiabatic fluctuations and flat
spatial geometry of inflationary models, a series of peaks are predicted
to occur in the CMBF data. The detailed predictions of the $\Lambda$CDM
and QCDM models are remarkably consistent with the observations.

To many in gravitational physics, however, the cosmological constant is
repulsive in more than one way. We should therefore welcome a scenario
proposed by Leonard Parker and Alpan Raval (PR) of the University of
Wisconsin at Milwaukee. The PR scenario agrees at least equally well with
both the SNe-Ia and CMBF data as the $\Lambda$CDM and QCDM models with no
more adjustable parameters than $\Lambda$CDM, need not be ``fine-tuned''
to have the desired properties, and is, in several ways, less {\it ad
hoc} than its competitors. They call their scenario the VCDM model since
the vacuum energy of a quantized scalar field provides negative pressure
to accelerate the Universe. 

The PR proposal, described in detail in [4, 5], adds a non-minimally
coupled ultra-low-mass free scalar field to (e.g.) the FRW-CDM model. The
required mass $m \approx 10^{-33}$ eV might be reasonable for a
pseudo-Nambu-Goldstone boson or even for the graviton. Given such a
scalar field, $\varphi$, standard techniques of quantum field theory in
curved spacetime may be used to construct the effective action
for the scalar field coupled to gravity by integrating out the
quantum fluctuations of the scalar field. PR discovered that it is
possible to perform a non-perturbative (i.e. infinite number of terms)
sum of all terms in the propagator with at least one factor of the scalar
curvature $R$. It is the non-perturbative effects which become dynamically
important on gigayear timescales when $R \approx m^2/(- \bar 
 \xi)$ where $\bar \xi = 0$ indicates conformal coupling and $\bar \xi =
1/6$ is minimal coupling and
$m$ is the mass of the scalar field. The single parameter $\bar m \equiv
m/\sqrt{\bar \xi}$ replaces $\Lambda$ in the VCDM scenario. In [5], PR
give a solution for the FRW scale factor $a(t)/a(t_j)$ for a VCDM model
containing vacuum energy, nonrelativistic matter, and radiation. It is
close to power law in $t$ for $t < t_j$ (as expected for zero
cosmological constant) and close to exponential in $t$ if $t > t_j$ (as
expected for non-zero cosmological constant). The non-perturbative vacuum
energy effects cause the transition at $t_j$ when the pressureless matter
density at $t_j$, $\rho_j = \bar m^2 / (8 \pi G)$. With this solution, it
is possible to construct the equation of state for the vacuum from the
Einstein tensor. The ultra-low-mass gives transition times corresponding
to cosmological redshifts of $z \approx 1$. 

Due to the mass scale, the non-perturbative effects are dynamically
negligible in the early universe.  (In [4c], PR point out that particle
creation---which is part of the non-perturbative effective action---might
be used to solve some current problems with the inflationary scenario.)
At late times, with the transition time controlled by the mass of the
scalar field, the vacuum energy dominates the dynamics. Vacuum energy
typically violates the energy conditions---in this case with a negative
pressure (but positive energy density). The FRW scale factor responds as
if there were a cosmological constant. Since $t_j \approx H_0^{-1}/2$ for
$H_0$ the present value of the Hubble parameter, the effective
cosmological constant turns on very late in the history of the universe.

Things to note about the PR scenario compared to the competition are that
given the existence of such an ultra-low-mass scalar field, there are no
more {\it ad hoc} assumptions. The VCDM equation of state results
from a wide range of values of $\bar m$. The current fits to SNe-Ia and
CMBF data by PR use preferred values of cosmological parameters
($\Omega_{\rm CDM} = 0.50$, $\Omega_{\rm B} = 0.06$, and $h = 0.7$ for
the CDM and baryon fractions and the current Hubble parameter in units of
100 km/s/Mpc). Should these change, it is still probable that a value of
$\bar m$ can be found to fit the data.

PR compare their VCDM model to a $\Lambda$CDM model with the same
cosmological parameters. The fits to the CMPF data are almost
indistinguishable. This is not true for the SNe-Ia data which has
information only back to $z \approx 1$. The VCDM model predicts
significantly fainter supernovae near $z \approx 1$ than does the
$\Lambda$CDM model. The VCDM prediction seems to follow the trend
in the data more closely. However, the current quality of the data does
not allow either model to be ruled out. Improved data might be able to
choose between these models.

Caldwell has recently argued [6] that the key ingredient in the
dependence of luminosity on redshift is $w$, the ratio of vacuum pressure
to vacuum energy density. Quintessence models appear to require $-1 < w <
0$ while PR's VCDM scenario yields $w < -1$. Caldwell demonstrated that
the latter behavior yields better fits to the SNe-Ia data but could only
generate a very contrived model with that property. The PR scenario
yields this behavior naturally.

Detailed properties of the PR scalar field scenario may be found in [4]
and references therein. Graphs showing the comparisons to the SNe-Ia and
CMBF data may be found in a very recent {\it Physical Review Letter} [5].
An even more recent discussion of this work as a plausible explanation of
the accelerating universe may be found in {\it Nature Science Update} [7].

Leonard Parker and Alpan Raval have discovered a scenario which might
represent the first observation of an effect predicted using quantum field
theory in curved spacetime as well as a new quantized field. In their own
words [5]: ``If the universe is indeed acting, through its own
acceleration, as a detector of this very low mass quantized field, then
there would be a wealth of implications for particle physics and
cosmology.'' 

{\bf References:}

[1] S. Perlmutter {\it et al.}, Nature {\bf 391}, 51 (1998); A. Riess {\it
et al.}, Astron. J. {\bf 116} 1009 (1998).

[2] R.R. Caldwell {\it et al.}, Phys. Rev. Lett. {\bf 80}, 1582 (1998).

[3] P. de Bernardis {\it et al.}, Nature {\bf 404}, 955 (2000); S. Hanany
{\it et al.}, Astrophys. J. Lett. {\it 545}, 5 (2000).

[4] L. Parker and A. Raval, Phys. Rev. D {\bf 60}, 063512 (1999);  {\bf
60}, 123502 (1999); {\bf 62}, 083503 (2000).  

[5] L. Parker and A. Raval, Phys. Rev. Lett. {\bf 86}, 749 (2001). 

[6] R.R. Caldwell, 
\htmladdnormallink{astro-ph/9908168}
{http://xxx.lanl.gov/abs/astro-ph/9908168}.

[7] Philip Ball, Nature Science Update,
\htmladdnormallink{http://www.nature.com/nsu/010208/010208-2.html}
{http://www.nature.com/nsu/010208/010208-2.html}.

\vfill\eject
\section*{\centerline {
The Lazarus Project:}\\
\centerline{Numerical relativity meets perturbation theory}}
\addcontentsline{toc}{subsubsection}{\it  
The Lazarus Project, by Richard Price}
\begin{center}
Richard Price, University of Utah
\htmladdnormallink{rprice@physics.utah.edu}
{mailto:rprice@physics.utah.edu}
\end{center}

Numerical relativity and gravitational wave detection exist in an
entangled state. Though there are many opinions about this state, two
things are probably not controversial. First, numerical relativity is
required to compute the dynamics and gravitational radiation when
inspiralling black holes merge, and second, that this is
extraordinarily difficult.  The difficulties prevent us from getting
answers to questions that are not only crucial to determining the
detectability of black hole mergers, but that are just plain
interesting.  Among these is the question of what happens when the
binary pair, late in its inspiral, has too much total (spin plus
orbital) angular momentum to form a Kerr hole. Does the inspiral
stall?  For inspiralling holes, is there a plunge or a gradual
transition from slow inspiral to distorted final hole? Does the answer
to this depend on such details as the spins?
In the past year or two there has been a slow but steady advance of the
frontier of the numerical relativity of binary black holes. True 3
dimensional runs have been successfully carried out for so-called
grazing collisions [1,2]  that describe
non-axisymmetric collisions of two holes that start fairly close
together with a fairly small impact parameter. Work is progressing in
many centers of numerical relativity on a better understanding of the
the outer boundary condition, how to excise the black hole (or avoid
the need for excision), how to choose the numerical variables for
greatest stability and/or accuracy, and much much more. This inspires
confidence that it will not be too long before numerical relativity
will be giving answers to astrophysical questions. 

That confidence has taken a nice jump in the past few months. A group
at the Albert Einstein-Max Planck Institute (``AEI'') has taken an
eclectic approach, called the Lazarus Project [3] to looking at black
hole mergers, and has provided waveforms generated by motion and
merger after the holes move inward from the ``ISCO'' (the Innermost
Stable Circular Orbit). The underlying idea is simple: If
straightforward numerical relativity is used, the code evolving the
spacetime would become unstable before useful information could be
extracted.  The Lazarus group therefore only uses numerical relativity
where it is indispensable: to evolve the gravitational field from the
ISCO stage to the point at which an almost stationary Kerr horizon is
formed. After this, relatively simple black hole perturbation theory
is used to continue the evolution of the spacetime. Perturbation
theory allows evolution to rise from its unstable grave and to
live again, like the biblical Lazarus. Sort of.

The initial Lazarus idea was the work of Carlos Lousto, Manuela
Campanelli, and John Baker, who were joined early in the project by
Bernd Br\"ugmann. A student, Ryoji Takahashi, has also recently been
added to the team. But the ``team'' in fact is the whole numerical
relativity group at the AEI, since the numerical relativity code of
the AEI and the CACTUS numerical relativity toolkit [4] constitute the
front end of the eclectic Lazarus approach. The back end is the code
to evolve perturbations of Kerr holes [5]  developed several
years ago. Not only did those elements already exist, but the idea
itself of doing late stage evolution with perturbation theory is not
new. The basic scheme had been set down in the mid 90s [6]
and applied to simple axisymmetric processes [7].  What the
Lazarus group did was to provide the tools for matching in a full 3D
problem in which the final hole is a rapidly rotating Kerr hole. The
details and difficulty of that task were what made this ``obvious''
step a real achievement.

The essence of the problem is to assign spacetime coordinates that are
in some sense almost Boyer-Lindquist coordinates for the numerically
evolved spacetime that is almost the Kerr spacetime outside the
horizon, and in those coordinates to identify perturbations. There is,
of course, no unique way of assigning coordinates and extracting
perturbations. Small changes in the coordinates induce small changes
in the perturbations. But the Teukolsky function, the quantity used
in Lazarus perturbative evolution, is changed only to second order
when small coordinate changes are made so the extraction process is
insensitive if the spacetime geometry is really ``almost Kerr.'' The
best way of building confidence about the method is inherent in the
method. The matching of numerical relativity and perturbation theory
is done at some transition time (i.e., on some spacelike
hypersurface).  If that transition time is taken to be too early, the
numerically evolved spacetime will not have achieved the point of
being a perturbation of Kerr.  If that transition time is too late the
numerical code will become unstable and the perturbations that are
extracted will be unrelated to the physical problem. If there is a
reasonable range of transition times that are neither too early nor
too late, then the Lazarus method should give the same results for all
transitions within this range. If the results are the same for a wide
range of transition times, then it is very hard to resist  accepting
the results.

During development, the Lazarus method passed many tests. Among them was a
test that results varied little for a range of transition times. But these
were results for small transverse momentum, and hence were grazing
collisions.  The big question is whether numerical relativity plus Lazarus
is ready to tell us about mergers, and the answer is either ``yes'' or
``very close.''  According to recent estimates [8,9], for puncture
type initial data [10], the ISCO for equal mass, nonspinning
holes, corresponds to a separation L of 4.8M (where M is the total ADM
mass), a transverse momentum for each hole of 0.335M, and total
angular momentum of $0.76M^2$.  These initial data were used as the
starting point for numerical evolution with an AEI code that uses
maximal slicing, zero shift, and the standard ADM approach. The range
of acceptable transition times was narrow, from about 10M to 12M, and
the radiated energy generated was uncertain by a factor of around
2. It would be nice, of course, to hear of 10\% accuracy, but a factor
of 2 uncertainty is really quite respectable.  Even with the limited
accuracy the results carry some important astrophysical
information. For one thing, there is an estimate of radiated energy
4-5\%$Mc^2$, more than twice as large as any previous computation
(though smaller than previous speculations), and a good omen for the
detectibility of black hole mergers, and for the yet larger numbers
that mergers of spinning holes may produce.)  Aside from a welcome
number, the results have an interesting qualitative lesson. There
appears to be little radiation associated with the early motion of the
holes; almost all the radiation can be ascribed to the  dynamics of
the distorted black hole that is formed.

It is possible, of course, that initial conditions that truly
represent the ISCO will tell a rather different story (such as more
orbital motion before the formation of a distorted horizon). To
explore such questions what is needed is the ability to start
evolution at an earlier time (or larger transverse momentum). This
cannot be done at present; the numerical code used for evolution would
go unstable before a perturbed Kerr hole is formed.  But this will
change, and that is perhaps the most exciting thing about the the
addition of Lazarus to the set of numerical relativity toos.  With
Lazarus, the improvements in code stability that will be achieved in
the coming months can quickly be turned into improvements in our
understanding of black hole mergers.
\vfill\eject

{\bf References:}

[1] S. Brandt {\em et al.}, Phys.Rev.Lett. {\bf 85} 5496 (2000).
[2] M. Alcubierre {\em et al.}, preprint  
\htmladdnormallink{gr-qc/0012079}{http://xxx.lanl.gov/abs/gr-qc/0012079}.

[3] J. Baker, B. Br\"ugmann, M. Campanelli, C. O. Lousto, and R. Takahashi,
\htmladdnormallink{gr-qc/0102037}{http://xxx.lanl.gov/abs/gr-qc/0102037};
 J. Baker, B. Bru\"ugmann, M. Campanelli, and
C. O. Lousto, Class. Quant. Grav. {\bf17}, L149
(2000). 
\htmladdnormallink
{http://www.aei-potsdam.mpg.de/\~{}lousto/lazaro.html}
{http://www.aei-potsdam.mpg.de/~lousto/lazaro.html}

[4] \htmladdnormallink
{http://www.cactuscode.org}{http://www.cactuscode.org}

[5] W. Krivan, P. Laguna, P. Papadopoulos, and N. Andersson, Phys.
Rev.~{\bf D56}m 3395 (1997). 

[6]  A. Abrahams and R. H. Price
Phys. Rev. D{\bf 53}, 1963 (1996).

[7] A. M. Abrahams, S. L. Shapiro and S. A. Teukolsky,
Phys.Rev. {\bf D51}, 4295(1995)

[8] G. B. Cook, Phys. Rev.~ {\bf D 50}, 5025 (1994).

[9] T. W. Baumgarte, Phys. Rev.~{\bf D62}, 024018 (2000). 

[10] S. Brandt and B. Br\"ugmann, Phys. Rev.~Lett. {\bf78}, 3606 (1997).

\vfill\eject
\section*{\centerline {
LIGO locks its first detector!}}
\addcontentsline{toc}{subsubsection}{\it  
LIGO locks its first detector!, by Stan Whitcomb}
\begin{center}
Stan Whitcomb, LIGO-Caltech
\htmladdnormallink{stan@ligo.caltech.edu}
{mailto:stan@ligo.caltech.edu}
\end{center}

In a year filled with important milestones, perhaps the most exciting event
for LIGO in 2000 was achieving first lock on its 2 km interferometer at the
Hanford Observatory.  To achieve the sensitivity required for searching for
gravitational waves the mirrors in the LIGO interferometers must be held in
the correct positions for the laser light to resonate properly, a process
known as "locking the interferometer". The required accuracies range from
0.1 to 100 picometers depending on which optic, a formidable task when one
realizes that the ground is continually in motion with an rms displacement
of about 1 micron!  Over the past year, we have been building toward the
goal of locking the full interferometer by locking simpler configurations
(for example a single arm cavity last spring) to test and tune the
photodetectors, electronics and software that are the heart of the locking
system.  Finally, last fall we began attempting to lock the full
interferometer.  To make the task a little easier in the beginning, we
introduced some additional losses into the optical path, either by
misaligning the optics slightly or by using a gate valve to clip the beams
in the arms.  After successfully locking the interferometer with the added
losses, we used the
locked sessions to characterize the servos and finally in January we were
able to lock the interferometer without the aid of the added losses.  This
marks an important transition in our commissioning effort--the beginning of
working with a fully functioning interferometer, and not just subsets of
the full instrument
(
\htmladdnormallink{\protect{\tt 
{http://www.ligo.caltech.edu/LIGO{\_}web/firstlock/}}}
{http://www.ligo.caltech.edu/LIGO\_{}web/firstlock/}).

Just a couple weeks after the final step in locking the 2 km
interferometer, we achieved another important milestone, this time on the
interferometer at the Livingston Observatory--the first 4 km arm cavity has
been locked to the laser.  This initiates the debugging and commissioning
of the control systems for lengths and angles on the second interferometer.

Each of LIGO's three interferometers has a well-defined role in the
commissioning.  The 2 km interferometer is the pathfinder--the place where
things are tried first and where problems are found.  The Livingston 4 km
interferometer is where systematic characterization and resolution of
problems takes place.  Installation of the Hanford 4 km interferometer
(still on-going) is paced so that it takes maximum advantage of the
experience with the other two interferometers, but is still completed on
schedule for the Science Run.

To help the LIGO Laboratory staff move toward round the clock operation and
to help the LIGO Scientific Collaboration (LSC) prepare to analyze the LIGO
data, we are carrying out a series of  "engineering runs". These runs have
durations of a few days to a few weeks and can involve either or  both
sites, depending on the goals of a particular run. The interferometers and
other measurement equipment are operated in a well-defined configuration,
and the data taken are archived and made available to LSC members.  In
early November 2000, the second of these engineering runs was held, with
the 2 km interferometer operating as a recombined Michelson interferometer
with Fabry-Perot arms (but no recycling).  The high percentage of time in
lock, about 90
third engineering run is scheduled for March.  The data from these
engineering runs both help us improve the detectors and prepare for the
task of analyzing the full LIGO data to come.

The remainder of 2001 has a busy schedule, bringing the three
interferometers into full operation at their design
sensitivity.  Everything is on track for the initiation of the LIGO Science
Run in early 2002.

\vfill\eject

\section*{\centerline {
Progress on the nonlinear r-mode problem}}
\addcontentsline{toc}{subsubsection}{\it  
Progress on the nonlinear r-mode problem, by Keith Lockitch}
\begin{center}
Keith Lockitch, Penn State
\htmladdnormallink{lockitch@gravity.phys.psu.edu}
{mailto:lockitch@gravity.phys.psu.edu}
\end{center}

Since the recent discovery that the r-modes of rotating stars are
unstable to the emission of gravitational waves [1], much 
effort has been directed towards improving the physical models of the
r-mode instability.
In the last issue of {\it Matters of Gravity}, Nils Andersson gave
an update on some of this work [2] - reviewing such effects
as neutron star superfluidity, the nonlinear evolution of the r-modes,
the damping associated with the formation of the crust and the
effects of general relativity on the spectrum and growth timescales
of the modes. (More detailed reviews may be found in [3].)
My purpose here is to report further on very recent progress
that has been made specifically on the nonlinear r-mode problem.

Early work suggested that the r-mode instability may limit the spin
rate of newly formed, rapidly rotating neutron stars and that the
radiation emitted while the star sheds its angular momentum may be
detectable by LIGO II [4].
The spin-down model on which these tantalizing estimates were based
assumed that the most unstable r-mode (with multipole indices $l=m=2$)
would be able to grow to an amplitude of order unity before being
saturated by some sort of nonlinear process.
It was also assumed that the star would spin down along a sequence
of stellar models each consisting of a uniformly rotating equilibrium
star perturbed by the dominant r-mode.

The central issue is whether the instability found in idealized 
models survives the physics that governs a young neutron star:
Will nonlinear coupling to other modes allow an unstable r-mode
to grow to unit amplitude? Does the background star retain a uniform 
rotation law as it spins down or does a growing r-mode generate 
significant differential rotation?
The importance of this last question was emphasized by Spruit 
[5] and by Rezzolla, Lamb and Shapiro [6] who argued 
that  differential rotation would wind up a toroidal magnetic field 
and drain the oscillation energy of the r-mode.
A number of different approaches have since been applied to the
nonlinear r-mode problem in an attempt to address these questions.

One notable approach is the direct numerical evolution of the
nonlinear equations describing a self-gravitating fluid.  Stergioulas
and Font [7] have performed 3-D general relativistic hydrodynamic
evolutions in the Cowling approximation, and Lindblom, Tohline
and Vallisneri [8] have performed 3-D Newtonian hydrodynamic
evolutions with an added driving force representing gravitational
radiation-reaction.

Stergioulas and Font [7] construct an equilibrium model of a
rapidly rotating relativistic star and add to it an initial
perturbation that roughly approximates its $l=m=2$ r-mode.
They then evolve the perturbed star using the nonlinear hydrodynamic
equations with the spacetime metric held fixed to its equilibrium value
(the relativistic Cowling approximation).
They find no evidence for suppression of the mode on a dynamical
timescale, even when the mode amplitude, $\alpha$, is initially 
taken to be of order unity.
Because of the approximate nature of the initial perturbation, other
oscillation modes are excited in the initial data. For a star with a
barotropic equation of state, the generic rotationally restored
mode is not a pure axial-parity r-mode, but an r-g ``hybrid'' mode
with a mixture of axial and polar parity components [8].
Stergioulas and Font [7] find that a number of these hybrid 
modes are excited in their initial data with good agreement between 
the inferred frequencies and earlier results from linear perturbation
theory [8]. In their published work, they find no evidence 
that the dominant mode is leaking its oscillation energy to other 
modes on a dynamical timescale.  Instead, a nonlinear version of an 
r-mode appears to persist over the time of the run, about 25 rotations
of the star. In additional runs with amplitudes substantially larger 
than unity, however, one no longer sees a coherent r-mode.  This may 
be evidence of nonlinear saturation, but further runs with more accurate 
initial data will be necessary to conclude this definitively [10].

These conclusions are consistent with preliminary results from studies
of nonlinear mode-mode couplings at higher order in perturbation
theory [11,12].
Other r-modes of a nonbarotropic star seem to give no
indication of a strong coupling to the $l=m=2$ r-mode unless its
amplitude is unphysically large ($\alpha
{\mathrel{\hbox{\rlap{\hbox
{\lower4pt\hbox{$\sim$}}}\hbox{$>$}}}} 30$!) [12].  Work is
still in progress on the nonlinear coupling of the dominant r-mode to
the g-modes of nonbarotropic stars [12] and to the hybrid modes of
barotropic stars [11].

The results of Stergioulas and Font [7] have also been confirmed and
significantly extended by the calculation of Lindblom, Tohline and
Vallisneri [8]. In Stergioulas and Font's calculation the growth 
of the unstable r-mode does not occur because the spacetime dynamics have
been turned off.  However, it would be impossible to model this growth
anyway even in a fully general relativistic hydrodynamic evolution,
because the timescale on which the mode grows due to the emission of
gravitational waves far exceeds the dynamical timescale of a rapidly
rotating neutron star.

To simulate the growth of the dominant r-mode in a calculation
accessible to current supercomputers, Lindblom, Tohline and Vallisneri
[8] take a different approach. They begin by constructing an
equilibrium model of a rapidly rotating Newtonian star and add to
it a small initial perturbation corresponding to its $l=m=2$ r-mode.
They then evolve the perturbed star by the equations of Newtonian
hydrodynamics with a post-Newtonian radiation-reaction force that
drives the current quadrupole associated with the $l=m=2$ r-mode.

By artificially scaling up the strength of the driving force, they
are able to shorten the growth time of the unstable r-mode by a factor
of $4500$.  In the resulting simulation the mode grows exponentially
from an amplitude $\alpha=0.1$ to $\alpha=2.0$ in only about 20
rotations of the star.

With this magnified radiation-reaction force, Lindblom, Tohline and 
Vallisneri [8] are able to confirm the general features of the
simplified r-mode spin-down models [4]. In their simulation, 
the star begins to spin down noticeably when the amplitude of the dominant
mode is of order unity, and ultimately about $40\%$ of the star's
angular momentum is radiated away. The evolution of the star's
angular momentum as computed numerically agrees well with the predicted
angular momentum loss to gravitational radiation.
If their model is accurate, however, gravitational radiation would 
not be emitted steadily at a saturation amplitude, but would die out 
after saturation and then reappear as the mode regenerates.

Again, there is no evidence of nonlinear saturation for mode amplitudes
$\alpha
{\mathrel{\hbox{\rlap{\hbox
{\lower4pt\hbox{$\sim$}}}\hbox{$<$}}}}
 1$.  The growth of the mode is eventually suppressed
at an amplitude $\alpha\simeq 3.4$, and the amplitude drops off sharply
thereafter.  Lindblom, Tohline and Vallisneri argue that the mechanism
suppressing the mode is the formation of shocks associated with the
breaking of surface waves on the star.  They find no evidence of
mass-shedding, nor of coupling of the dominant mode to the other
r-modes or hybrid modes of their Newtonian barotropic model.

These various studies all provide evidence pointing to the same
conclusion: the most unstable r-mode appears likely to grow to an
amplitude of order unity before being suppressed by nonlinear
hydrodynamic processes. It is important to emphasize, however,
that the 3-D numerical simulations have probed nonlinear processes
occurring only on dynamical timescales and that the actual growth
timescale for the r-mode instability is longer by a factor of order $10^4$.
It is possible that the instability may be suppressed by hydrodynamic
couplings occurring on timescales that are longer than the dynamical
timescale but shorter than the r-mode growth timescale.
Further work clearly needs to be done before definitive conclusions
can be drawn. Particularly relevant will be the results from the
ongoing mode-mode coupling studies [11,12].

Turning to the question of differential rotation, deviations from
a uniform rotation law are observed in both of the 3-D numerical
simulations [7,8]
It has been proposed that differential rotation will be driven by
gravitational radiation-reaction [5] as well as being associated
with the second order motion of the r-mode, itself [6].
In a useful toy model, Levin and Ushomirsky [13] calculated an
exact r-mode solution in a thin fluid shell and found both sources
of differential rotation to be present.

To address in more detail the issue of whether or not the r-mode
instability would generate significant differential rotation,
Friedman, Lockitch and S\'a [14] have calculated the 
axisymmetric part of the second order r-mode.
We work to second order in perturbation theory with the equilibrium
solution taken to be either a slowly rotating polytrope (with index
$n=1$) or an arbitrarily rotating uniform density star (a Maclaurin
spheroid).  The first order solution, which appears in the source
term of the second order equations, is taken to be a pure $l=m$
r-mode with amplitude $\alpha$.

We find that differential rotation is indeed generated both by
gravitational radiation-reaction and by the quadratic source terms
in Euler's equation; however, the latter dominate a post-Newtonian
expansion.
The functional form of the differential rotation is independent
of the equation of state - the axisymmetric, second order change
in $v^\varphi$ being proportional to $z^2$ (in cylindrical coordinates)
for both the polytrope and Maclaurin.

Our result extends that of Rezzolla, Lamb and Shapiro [6]
who computed the order $\alpha^2$ differential drift resulting from 
the linear r-mode velocity field.
These authors neglect the nonlinear terms in the fluid equations
and argue (based on an analogy with shallow water waves) that the
contribution from the neglected terms might be irrelevant.  Indeed,
for sound waves and shallow water waves, the fluid drift computed using 
the linear velocity field turns out to be exact to second order 
[15]; thus, one may safely ignore the nonlinear terms.
However, for the motion of a fluid element associated with the r-modes, 
we find that there is in fact a non-negligible contribution from the 
second-order change in $v^\varphi$. Interestingly, the resulting second 
order differential rotation is stratified on cylinders.
It remains to be seen whether the coupling of this differential rotation 
to the star's magnetic field does indeed imply suppression of the r-mode 
instability.

{\bf References:}

[1]
Andersson, N., Astrophys. J., {\bf 502}, 708, (1998);\\
Friedman, J. L. and Morsink, S. M., Astrophys. J., {\bf 502}, 
	714, (1998)

[2]
Andersson, N., {\it An update on the r-mode instability}, MOG No. 16, 
(2000)\\ 
\htmladdnormallink{http://gravity.phys.psu.edu/mog.html}
{http://gravity.phys.psu.edu/mog.html}

[3]
Friedman, J. L. and Lockitch, K. H., Prog. Theor. Phys. Supp., 
{\bf 136}, 121 (1999);
Andersson, N. and Kokkotas, K. D., {\it The r-mode instability
in rotating neutron stars}, preprint 
\htmladdnormallink{gr-qc/0010102}{http://xxx.lanl.gov/abs/gr-qc/0010102};
Lindblom, L., {\it Neutron star pulsations and instabilities},
preprint 
\htmladdnormallink{astro-ph/0101136}{http://xxx.lanl.gov/abs/astro-ph/0101136}

[4]
Lindblom, L., Owen, B. J. and Morsink, S. M., Phys. Rev. Lett.,
{\bf 80}, 4843, (1998);
Andersson, N., Kokkotas, K. and Schutz B. F., Astrophys. J., 
{\bf 510}, 846, (1999);
Owen, B. J., Lindblom, L., Cutler, C., Schutz, B. F., Vecchio, A. and
Andersson, N., Phys. Rev. D, {\bf 58}, 084020, (1998)

[5]
Spruit, H. C., Astron. and Astrophys., {\bf 341}, L1, (1999)

[6]
Rezzolla, L., Lamb, F.K. and Shapiro, S.L., Astrophys. J. Lett., 
{\bf 531}, L139, (2000)

[7]
Stergioulas N. and Font, J.A., {\it Nonlinear r-modes in rapidly
rotating relativistic stars}, Phys. Rev. Lett., in press (2001);
preprint 
\htmladdnormallink{gr-qc/0007086}{http://xxx.lanl.gov/abs/gr-qc/0007086}

[8]
Lindblom, L., Tohline, J. E. and Vallisneri, M., {\it Non-linear evolution
of the r-modes in neutron stars}, Phys. Rev. Lett., in press (2001); 
preprint 
\htmladdnormallink{astro-ph/0010653}{http://xxx.lanl.gov/abs/astro-ph/0010653}

[9]
Lockitch, K. H. and Friedman, J. L., Astrophys. J., {\bf 521}, 764, (1999);
Lockitch, K. H., Andersson, N. and Friedman, J. L., Phys. Rev. D, {\bf 63},
024019, (2000)

[10]
Stergioulas, N., private communication (2001).

[11]
Schenk, A. K., Arras, P., Flanagan, \'E. \'E., Teukolsky, S. A. and
Wasserman, I., {\it Nonlinear mode coupling in rotating stars and the
r-mode instability in neutron stars}, preprint 
\htmladdnormallink{gr-qc/0101092}{http://xxx.lanl.gov/abs/gr-qc/0101092}

[12]
Morsink, S. M., private communication, (2000)

[13]
Levin, Y. and Ushomirsky, G., preprint 
\htmladdnormallink{astro-ph/0006028}{http://xxx.lanl.gov/abs/astro-ph/0006028}, (2000)

[14]
Friedman, J. L., Lockitch, K. H. and S\'a, P. M., in preparation (2001)

[15]
Lamb, F. K., Markovi\'c, D., Rezzolla, L. and Shapiro, S. L., 
private communication (1999)

\vfill\eject

\section*{\centerline {
Analog Models of General Relativity}}
\addtocontents{toc}{\protect\medskip}
\addtocontents{toc}{\bf Conference Reports}
\addtocontents{toc}{\protect\medskip}
\addcontentsline{toc}{subsubsection}{\it  
Analog Models of General Relativity, by Matt Visser}
\begin{center}
Matt Visser, Washington University St. Louis
\htmladdnormallink{visser@kiwi.wustl.edu}
{mailto:visser@kiwi.wustl.edu}
\end{center}

The workshop ``Analog Models of General Relativity'' was held in Rio
de Janeiro from 16 October to 20 October 2000. The organizing
committee consisted of Mario Novello, Grigori Volovik, and
myself. Invited speakers talked about a wide range of issues
concerning the use of condensed matter systems as analogues of (and
analogs for) general relativity. Condensed matter analogs can be used
to help us understand GR, or GR can be used to help us understand
condensed matter physics. More boldly, you can use condensed matter
analogs to suggest possible replacements for GR, physical systems that
approximate ordinary GR in the appropriate limit. Among the invited
presentations:

(1) The workshop started with an introductory survey, presented by
myself, that set the basic parameters for the week.

(2) Bill Unruh talked about his acoustic black holes (``dumb holes''),
illustrating the way that acoustics in a moving fluid leads to the
notion of an ``effective acoustic metric''.

(3) Grigori Volovik discussed the use of $^3{}He$ as a model of, and
indeed for, GR. (Low-energy quasiparticles near the Fermi surface can
in certain circumstances generically exhibit a relativistic spectrum,
and induced gravity a la Sakharov can then be argued to lead to an
effective dynamics similar to Einstein gravity.)

(4) Brandon Carter carefully distinguished the notions of
quasi-gravity from pseudo-gravity. (q-gravity: systems that
mathematically simulate GR but are qualitatively different,
e.g. acoustic geometries; p-gravity: systems that physically mimic
gravitational fields, e.g. centrifugal force). He also discussed a
model of how to use braneworld cosmologies to mimic gravity in a
non-standard way.

(5) Ulf Leonhardt described his proposal for an ``optical horizon''
using ``slow light'' (resonance induced transparency in a
Bose--Einstein condensate; a BEC).

(6) Renaud Parentani talked about quantum metric fluctuations and
Hawking radiation. He argued that the near horizon propagation of
outgoing quanta resembles that of photons in a moving random medium.

(7) Haret Rosu discussed a number of topics concerning exotic effects
in the GR quantum interface.

(8) Mario Novello described nonlinear electrodynamics (for example,
Born-Infeld, Schwinger, or Euler-Heisenberg electrodynamics) and the
way it leads to the notion of an effective metric governing photon
propagation.

(9) Ted Jacobson talked about a particular implementation of the
notion of ``analog horizon'' in a $^3{}He$ superfluid system.

(10) Mike Stone presented a careful discussion of how notions of
effective metric and the machinery of general relativity can help
understand the concepts of pseudo-momentum and physical momentum in
condensed matter systems.

In addition there were a number of contributed talks (Santiago
Bergliaffa presented examples of gravity-like systems in non-linear
electrodynamics, Jose Salim discussed closed spacelike photon paths
in nonlinear electrodynamics, Nami Fux Svaiter discussed the rotating
vacuum and the quantum Mach principle, and Carlos Barcelo presented a
discussion of analog gravity based on Bose--Einstein condensates; BECs).

Additionally, approximately 50 graduate students and postdocs attended
the workshop.

There was considerable animated discussion, aided by an open workshop
format that left plenty of time for give-and-take. The most promising
systems for experimentally mimicking ``event horizons'' seem to be
based on (a) ``slow light'' in BEC systems with resonance induced
transparency, (b) quasiparticles in superfluids, and (c) acoustic
oscillations of the phase of the condensate field in BECs.
Considerable enthusiasm and hope was expressed that one or more of
these ``analog systems'' might be brought to laboratory fruition in
the near (5 to 10 year) future.

Many of the transparencies from the presentations (plus some
write-ups, web-links, and other technical information) is now
available from the post-conference website:
\htmladdnormallink{{\tt
http://www.lafex.cbpf.br/\~{}bscg/analog/}}
{http://www.lafex.cbpf.br/\~{}bscg/analog/}

A mirror of this website is maintained in the USA at:
\htmladdnormallink{{\tt
http://www.physics.wustl.edu/\~{}visser/Analog/}}
{http://www.physics.wustl.edu/\~{}visser/Analog/}
\vfill\eject
\section*{\centerline {
Workshop on Astrophysical Sources}\\
\centerline{of Gravitational
Radiation}\\\centerline{ for Ground-Based Detectors}}
\addcontentsline{toc}{subsubsection}{\it
Astrophysical Sources of Gravitational radiation, by Joan Centrella}
\begin{center}
    Joan Centrella, Drexel University
\htmladdnormallink{joan@sparrow.drexel.edu}
{mailto:joan@sparrow.drexel.edu}
\footnote{As of April 2, 2001, \htmladdnormallink
{jcentrel@lheapop.gsfc.nasa.gov}
{mailto:jcentrel@lheapop.gsfc.nasa.gov}}
\end{center}

As the $21^{\rm st}$ century begins, gravitational wave
astronomy is poised for unprecedented expansion and discovery.
Understanding the
expected gravitational wave frequencies and other
characteristics of astrophysical sources is essential to take
full advantage of these opportunities, and to stimulate
and influence detector development.  To this end, gravitational
wave experimentalists, relativists, astronomers, and astrophysicists
met at Drexel University on October 30 - November 1, 2000 for a
workshop focusing on gravitational wave sources for ground-based detectors.

The scientific sessions began with a series of talks on the detectors.
Barry
Barish presented a review of first generation interferometers.  He was
followed by Peter Fritschel, who described the current plans for
LIGO-II, and Kip Thorne, who discussed issues involving thermal noise,
optical noise, and quantum non-demolition for instruments beyond
LIGO-II.
Bill Hamilton then gave an overview of resonant bar detectors and the
international bar detector community.

The data expected
from the LIGO-I science run was addressed by Albert Lazzarini,
who also discussed the GriPhyN project and its relevance to LIGO
data. Patrick Brady discussed LIGO data analysis efforts,
and  Sam Finn followed
with a description of LIGO's science reach.

New initiatives in astronomy and astrophysics provide rich resources
and partnerships for gravitational wave astronomy.  Tom Gaisser reported
on the recommendations of the Particle, Nuclear, and Gravitational
Wave
Astrophysics panel from the recently completed decadal survey
 {\em Astronomy and
 Astrophysics in the New Millennium}
(see, e.g., 
\htmladdnormallink
{{\tt http://www.nap.edu/books/0309070317/html/}}
{http://www.nap.edu/books/0309070317/html/}
).  He was
followed by Tom Prince, who described the space-based LISA detector,
and Bob Hanisch, who discussed the National Virtual Observatory; both
of these projects received strong support from the decadal survey.
Nick White gave an overview of NASA's future programs in high energy
astrophysics, many of which focus on black holes and their
environments.

Coalescing compact binaries constitute the
``bread and butter'' source for ground-based interferometers, and
were addressed from a variety of directions.  Vicky Kalogera began
with a discussion of event rates for binary inspiral; she was followed
by Steve McMillan, who described the formation of black hole binaries
in globular clusters.  The importance of large scale numerical
simulations was highlighted by numerous speakers.  Josh Faber and Fred
Rasio presented new work on the hydrodynamics of neutron star mergers,
and Max Ruffert underscored the importance of coalescing compact
binaries for understanding gamma ray bursts.  William Lee presented
simulations of black hole-neutron star coalescence, and Richard
Matzner discussed binary black hole collisions.  Thomas Baumgarte
concluded this session with a talk on the innermost stable circular
orbit in compact binary systems.

Cosmological sources of stochastic gravitational waves were addressed
by Arthur Kosowsky.  David Spergel spoke on plans to use the CMB as a
gravitational wave detector.

Stellar core collapse has long been proposed as a source of
gravitational waves, and was discussed by Chris Fryer.  Kimberly New
described dynamical rotational instabilities that can arise in
centrifugally hung compact cores, and David Brown discussed their
occurrence during collapse.  Gravitational radiation from secular
bar-mode instabilities was addressed by Dong Lai.

In recent years, a number of exciting new developments have arisen in
the study of rotating neutron stars.
  Jean Swank discussed X-ray observations of accretion
instabilities on long and short timescales in low mass X-ray binaries.
Tod Strohmayer
described X-ray observations giving evidence for millisecond spins.
Gravitational radiation produced by temperature gradients, and its
importance for LIGO-II, was addressed by Lars Bildsten.  Greg
Ushomirsky discussed gravitational waves from r-modes in accreting
neutron stars and young neutron stars.

Conference rapporteurs Rainer Weiss, Peter Saulson, and Joel Tohline
provided thought-provoking and insightful overviews of the meeting.

This workshop proved to be a fruitful and enjoyable time for these
different communities to interact with each other.  Those who were
unable to attend in person should visit
\htmladdnormallink
{{\tt http://www.physics.drexel.edu/events/astro{\_}conference}}
{http://www.physics.drexel.edu/events/astro{\_}conference}.  At this
website the transparencies of the talks can be viewed by going to the
meeting program, and clicking on the title of each talk.

\vfill\eject
\section*{\centerline {
Numerical Relativity and Black Hole Collisions }\\\centerline{
at the 20th Texas Symposium on Relativistic Astrophysics}}
\addcontentsline{toc}{subsubsection}{\it
Numerical relativity at the 20th Texas meeting, by Pablo Laguna}
\begin{center}
    Pablo Laguna, Penn State
\htmladdnormallink{pablo@astro.psu.edu}
{mailto:pablo@astro.psu.edu}
\end{center}

In spite of the low general attendance to the 
20th Texas Symposium on Relativistic Astrophysics last
December in Austin, Texas, the 
parallel session on numerical relativity and black hole collisions
was not only oversubscribed and had to be extended one hour
beyond the allocated time limit, but it also attracted a large audience.
Eleven ten minute talks were given and two one minute poster
advertisements.

Bernd Schmidt (AEI/Germany) presented results on the
numerical evolution of the Kruskal spacetime using the conformal field
equations. Specifically, he addressed
initial data sets for the conformal field equations which describe
spacelike hypersurfaces in the conformally extended Kruskal spacetime.
These are data sets that have been evolved using the code for the
conformal field equations developed by P. Huebner. Schmidt showed
results from these simulations.

Sascha Husa (AEI/Germany) reported recent progress toward the global
study of asymptotically flat spacetimes with numerical relativity.
The development of a 3D solver for asymptotically Minkowski extended
hyperboloidal initial data has rendered possible the application of
Friedrich's conformal field equations to astrophysically interesting
spacetimes. As a first application, he presented the future development
of a hyperboloidal set of weak initial data, including future null and timelike
infinity. Using this example, he sketched the numerical techniques
employed
and highlighted some of the unique capabilities of the numerical code.
Husa briefly mentioned the implications of these results
for future work on (multi) black hole spacetimes.

Pedro Marronetti (Texas/Austin) presented the first full numerical 
solutions of the initial data problem of two black 
holes based on a Kerr-Schild spacetime slicing.
These new solutions provides more physically realistic solutions
than the initial data based on conformally flat metric/maximal slicing
methods. The singularity/inner boundary problems are circumvented by a
new technique that allows the use of an elliptic solver on a Cartesian
grid where no points are excised, simplifying enormously the numerical
problem. After this presentation, Richard Matzner (Texas/Austin) showed a
video of the simulation of grazing collisions of black holes performed
by the Texas-Pittsburgh-Penn State collaboration.

Deirdre Shoemaker (Penn State)
gave a presentation pointing out first that
recent experience with numerically evolving the
space-time of grazing collisions of black holes have
provided valuable lessons about the difficulties that
one might face in the more important case of a collision
when the holes start far apart. She stressed that
some of the difficulties can be successfully modeled
and studied in attempting to understand how to evolve
a single black hole. She then presented results from
several studies that the Penn State group have
performed attempting to evolve the
ADM equations with various lapse/shift conditions for this
problem.

E. Seidel and R. Takahashi (AEI/Germany) presented results from
the full 3D evolution of two colliding black
holes, with angular momentum, spin, and unequal mass. They emphasized
that
the AEI group has for the first 
time computed waveforms a grazing collision.  The collision
can be followed through the merger to form a single black hole, and
through part of the ring-down period of the final black hole.  The
apparent horizons are tracked and studied, and physical parameters,
such as the mass of the final black hole, are computed.  The total 
energy radiated is shown to be  consistent with the total ADM mass of 
the spacetime and the final
black hole mass.  Finally, Seidel discussed the implications of 
these simulations for
gravitational wave astronomy.

Miguel Alcubierre and D. Pollney (AEI/Germany) discussed
a series of techniques required for
the numerical simulation of black hole spacetime.
These techniques include the choice of an adequate
formulation of the evolution equations, the choice
of lapse and shift conditions, the choice of boundary
conditions, and the use of black hole excision.
They presented also the results of the three dimensional
simulation of a distorted black hole where these
techniques have been applied successfully, allowing
us to obtain long-term stable, accurate simulations.

Carlos Lousto, John Baker and Manuela Campanelli (AEI/Germany)
presented results from 
the coalescence of binary black holes from the innermost stable
circular orbit down to the final single rotating black hole
under the Lazarus framework. The Lazarus approach 
combines the full numerical approach to solve Einstein
equations, applied in the truly nonlinear regime, and linearized
perturbation theory around the final distorted single black hole at later
times.  Their results indicate a significantly higher amount
of energy radiated (up to 4 or 5
They presented also waveforms lasting for over t~100M, 
and pointed out that their waveforms suggest an early
nonlinear ringing.

Peter Diener (AEI/Germany) presented results from
a work in progress on Binary black hole initial data,
based on adding two Schwarzschild black holes in Kerr-Schild form.
Using attenuation functions in order to force the constraint deviations
to vanish near the singularities, he pointed out that
it is possible to solve the constraint
equations over the entire spatial grid.

S. Hawley (AEI/Germany) presented results of his study on
critical phenomena in boson stars. Specifically,
this study introduces a real field 
to perturb the boson star via a gravitational interaction 
which results in a significant transfer of energy.
The resulting critical solutions not only 
are similar to those of unstable boson stars
but also persist for a finite time before dispersing or forming a black hole. 

John Whelan (Texas/Brownsville)
presented results in which the quasi-stationary approximation 
is used to model a phase of the
inspiral of a compact-object binary where the time scale for decay of
the orbits is long compared to the orbital period, without imposing
any weak-field approximations.  These results were obtained
by numerically solving for a
stationary spacetime which approximates the slowly evolving one,
maintaining equilibrium in the radiating system by imposing a balance
of incoming and outgoing radiation at large distances.  Such a
radiation-balanced solution can serve as an alternative to existing
techniques for constructing initial-value data for full-numerical
"plunge" simulations.

Mark Miller (WashU) presented a new method for numerically constructing
solutions 
to the constraint equations of general relativity that correspond to a
single black hole in quasi-circular orbit with a single neutron star.
By examining sequences of such solutions, he showed that it is possible to
estimate
the location of the innermost stable circular orbit for these systems,
which will be used as the starting point for full 3D numerical 
simulations of binary black hole - neutron star coalescences.  

Harald Dimmelmeier (Garching/Germany)
reported on results from a new numerical relativistic 
hydrodynamical code for axisymmetric core collapse. He utilizes 
high-resolution shock capturing methods for the hydrodynamic
equations, and Wilson and Mathews' approximation of a conformally
flat spatial metric. The results presented were 
obtained from simulations of supernova core 
collapse and bounce, and rotating neutron star simulations. 
He showed gravitational radiation waveforms obtained by 
post-processing using the quadrupole formula.

\end{document}